\RequirePackage{fix-cm}
\documentclass{svjour3}                     % onecolumn (standard format)
\smartqed  % flush right qed marks, e.g. at end of proof
\usepackage{graphicx}
\usepackage{wrapfig}
\usepackage{xcolor}
\usepackage{adjustbox}  
\begin{document}

\title{Compress-Store on Blockchain: A Decentralized Data Processing and Immutable Storage for Multimedia Streaming%\thanks{Grants or other notes
%about the article that should go on the front page should be
%placed here. General acknowledgments should be placed at the end of the article.}
}
%\subtitle{A Decentralized Data Processing and Immutable Storage for Multimedia Streaming}

%\titlerunning{Short form of title}        % if too long for running head

\author{Suayb S.~Arslan        \and
        Turguy Goker %etc.
}

%\authorrunning{Short form of author list} % if too long for running head

\institute{Suayb S. Arslan \at
              MEF University, Sarıyer, Istanbul, Turkey. \\
              Tel.: +90-212-395-3735\\
              \email{arslans@mef.edu.tr}           %  \\
%             \emph{Present address:} of F. Author  %  if needed
           \and
           Turguy goker \at
           Quantum Corporation, Irvine, CA, USA. 
}

\date{Received: date / Accepted: date}
% The correct dates will be entered by the editor

\maketitle

\begin{abstract}
Decentralization for data storage is a challenging problem for blockchain-based solutions as the blocksize plays the key role for scalability. In addition, specific requirements of multimedia data calls for various changes in the blockchain technology internals. Considering one of the most popular applications of secure multimedia streaming, i.e., video surveillance, it is not clear how to judiciously encode incentivization, immutability and compression into a viable ecosystem. In this study, we provide a genuine  scheme that achieves this encoding for a video surveillance application. The proposed scheme provides a novel integration of data compression, immutable off-chain data storage using a new consensus protocol namely, Proof-of-WorkStore (PoWS) in order to enable fully useful work to be performed by the miner nodes of the network. The proposed idea is the first step towards achieving greener application of blockchain-based environment to the video storage business that utilizes system resources efficiently.  
% \PACS{PACS code1 \and PACS code2 \and more}
% \subclass{MSC code1 \and MSC code2 \and more}
\end{abstract}

\section{Introduction}
\label{intro}

According to recent estimations, there will be over {50} billion connected devices by {2022}, all of
which will generate and then require management, storage, and retrieval of large size of data \cite{Nordrum2016}.
Connected devices, {which typically goes by the name Internet of Things (IoT),} combined with consumer-based applications and the increasing need to
share data across different business lines, are all playing their part in increasing demand for {effective/efficient}
processing and data storage. Some of these data inherently requires immutability and calls for long term retention. For instance, think about government archiving or another popular example of video/data surveillance.
Businesses desiring to launch new, data-driven applications are bound to confront with
incredible amount of time, effort and coordination to provision new databases today. Now we
begin to see dominant commercial and revenue dependency on data which leads to large
volumes being stored in vulnerable centralized databases (even in the cloud), creating privacy
and durability risks at a scale seldom seen before in history.

Today, the dominant practice in unstructured data storage is based on a local or remote single
system architecture or cloud-based file/block/object storages (such as Amozon S3 \cite{amazons3} etc.) which
are still highly centralized. Although they can be distributed, they are still in governance of a single body of management and hence these systems are definitely
considered as a beacon for hackers (both external and internal) looking to attack. They also have many points of failure
should the managing company's ecosystem is affected by an unpredictable system error or experiences down time
as a result of a power outage. In addition, data type being stored has an immense effect on the management decisions. For example, multimedia sources are time dependent series of data
and must carefully be protected and communicated by paying attention to streaming
requirements. In contrast, decentralized storage does not encounter these problems because it
utilizes geographically distributed anonymous or permitted individual nodes, either regionally or
globally. Hence, meeting point of any applications based on decentralized video involves several challenges to tackle. One of the proven paradigm for storage is known as Distributed Ledger Technology (DLT) \cite{pilkington2016}.

DLT can be implemented using different consensus algorithms to ascertain that the world
view of each node is the same. Old traditional way is centered around voting-based consensus
such as Paxos \cite{lamport2001}  then the more understandable version that goes with the name Raft \cite{raft2014}. Most recently, random {(probabilistic)} consensus algorithms have gained popularity.
One of the consensus approaches to DLT that became quite common in decentralized cryptocurrency
market is blockchain \cite{baliga2017}. For more details about the blockchain technology, future trends and challenges, please see \cite{mermer2018}. Considering some open-source public blockchains (such as
Bitcoin \cite{nakamoto2008} and Ethereum \cite{wood2014}), the set of transactions that are stored within the linked-list of blocks
generates a type of decentralized database or storage of structured data. However due to
scalability concerns, the size of blocks cannot grow very large and hence it is not hard to see
that these public blockchains are not designed for bulk data storage and management, and using them to do so would consume too much local space, too much time for processing and too much energy to fulfill all the executions. In fact, it has the potential to make the system centralized should the parties participating are in possession of extensive resources. 

{A distributed replicated database, which stores data that can be shared among all system participants, is one of Blockchain's most important applications. Such a framework can be a basis for storage in IoT systems as evidenced by recent works \cite{tseng2020}, \cite{arslaniot2020}. However, the data storage must be judiciously handled in the resource-constrained devices of future IoT applications. Before diving into the details of how the storage is handled in the proposed architecture}, let us explore some of the decentralized data storage {or database} options previously devised and implemented.

\begin{table*}[t!]
\makebox[\linewidth]{
\begin{tabular}{|l|l|l|l|l|l|}
\hline
\textbf{Project} & \textbf{\begin{tabular}[c]{@{}l@{}}Smart\\   Contracts\end{tabular}} & \textbf{\begin{tabular}[c]{@{}l@{}}Multi-region\\   redundancy\end{tabular}} & \textbf{Feature}                                                                     & \textbf{Consensus}     & \textbf{\begin{tabular}[c]{@{}l@{}}Scalability\\   (1-3)\end{tabular}} \\ \hline
Sia              & Yes                      & Yes                                                                          & \begin{tabular}[c]{@{}l@{}}Archieving, \\ Very decentralized \\ own BC\end{tabular}     & BFT                    & 2                          \\ \hline
Storj            & No                       & Yes                                                                          & \begin{tabular}[c]{@{}l@{}}Object,ECC encrypted,\\   sharded, DHT,  ETH\end{tabular} & \begin{tabular}[c]{@{}l@{}}Proof of\\   Retrievability\end{tabular}  & 1                          \\ \hline
ETH Swarm        & Yes                      & Yes                                                                          & DHT, ETH                                                                             & \begin{tabular}[c]{@{}l@{}}Proof of\\   Retrievability\end{tabular} & 1                          \\ \hline
FileCoin         & Yes                      & Yes                                                                          & IPFS, Replication                                                                    & \begin{tabular}[c]{@{}l@{}}Proof of\\   Replication\end{tabular}  & 1-2                        \\ \hline
MaidSafe         & No                       & Yes                                                                          & No Blockchain                                                                        & \begin{tabular}[c]{@{}l@{}}close Group\\   Consensus\end{tabular}  & 3                        \\ \hline
\end{tabular}
}
\vspace{2mm}
\caption{Some decentralized cloud data storage projects centered around distributed technologies. Provided is a rough and relative estimation of scalability using a range of 1-3. Larger the number is, better scalability it possesses. BC: Blockchain. BFT: Byzantine Fault Tolerance}
\vspace{-1mm}
\end{table*}

\subsection{\textit{Storing data on the blockchain:}} Blockchains are immutable constructs and hence do not
allow random access for write and frequent changes. Also, only limited number of blocks can be securely added to the chain for a given time period, which makes the throughput fail to meet most of the
data storage requirements. In addition, since the size of the data might be arbitrarily large and
\textit{full nodes} are supposed to store the entire blockchain, the capacity required  storing it will
eventually exceed the persistent storage space of many full nodes of the network \cite{Karafiloski2017}. Thus, only a
specific set of nodes in the network would be able to hold the entire blockchain. {This will result in a centralization problem, in which a small number of nodes will control the database system, resulting in a loss of security since only a few known and capable nodes will be able to actively engage in the mining process.}

\subsection{\textit{Peer-2-{P}eer file systems:}}  This approach is based on sharing files on client computers and
uniting them using a global file system interface. This technology utilizes a similar protocol to
BitTorrent \cite{bittorent2005} and  Distributed Hash Table (DHT) concepts. Unlike {access points such as} IPs and ports, the data contents will be content--addressable using hashes of the content allowing the separation of storage
location and data. Data is available only if the nodes storing the copies are online. Once the data
content is replicated enough number of times, the availability/reliability of data is no longer a 
concern. DHT-based technology serves only static files which can not be modified or removed
once uploaded. The deletion of files cannot be ensured as this technology is not intended to do
so. In other words, the number of copies are not determined by the system but rather the request
pattern on that data by the network nodes. Lastly, the stored files cannot be searched by their
meaningful content. One of the well known succesful  implementations of this idea is known as InterPlanetary File
System (IPFS) \cite{ipfs2004} {based on Kademlia DHT \cite{maymounkov2022}. In IPFS, objects are immutable, which means that new versions of an object contain different content and hence a different hash value than earlier ones. For more comprehensive survey about DHT-based architectures, please see \cite{nazadeh2021}.}

\subsection{\textit{Decentralized cloud file storages:}}  Most of the decentralized systems {share some commonalities with} centralized cloud file
storages such as Dropbox \cite{drago2012}. {In particular,} peers in the network offer their unused persistent storage space for
rent and gets rewards in return for providing data storage space and services. Some of the examples include
Sia \cite{vorick2014}, Storj \cite{storj2014}, Swarm \cite{hartman1999}, Filecoin \cite{filecoin2014} and MaidSafe \cite{maidsafe2014} which are listed and summarized in Table 1 based on
the technologies they are made of. These storage systems provide highly reliable, enormous
capacity with varying degrees of access latency and security. As can be seen most of them are based on a type of blockchain implementation and backed by some kind of incentivization mechanism. Thus, these projects are intended to serve static files only, and no content
search is allowed (unless a specific feature gets added as they are all evolving projects) and,
since they are built on peer’s or anonymous rented hardware, they are not free of charge. All these projects are
optimized for file storage (show decent performance with file accesses) but fairly fall short in
accommodating for time-series data (such as Multimedia or IoT data etc.) An example of such data include append-only data streams, with a single writer and lots of readers. Although recently few attempts are made towards creating archival data storage and sharing ecosystems for IoT systems, none adequately addresses the streaming data requirements \cite{shafagh2017}.

\subsection{\textit{Blockchain-based solutions for copyright protection and video hashing:}}  Blockchain technology is very attractive solution for online electronic notary services, document certification, proof of ownership and authenticity. Most of such initiatives targeted mobile devices and application development environments whereas the blockchain formed the back-end registrar for document hashes and related information etc. In some of these applications, decentralized database systems are preferred (such as BigChainDB \cite{bigchaindb2016} or TiesDB) and the rest use content-addressable decentralized options (such as IPFS). Examples include initiatives such as Block Notary, Stampery \cite{stampery2015}, Verif-y \cite{verify2018}. On the other hand, there are also available video sharing and video streaming services based on blockchains \cite{livepeer2017}. These services verify ownership of each video content as a whole. LIVEPEER for instance is structured for broadcasting by transcoding video source into all formats and bitrates. Flixxo and Viuly are video sharing platforms \cite{liu2016}, in a way competitor projects to Youtube Inc., by offering an entirely decentralized platform in which, contrary to their competitor cloud-based providers, not only content generators are rewarded but also are the content viewers as well. Viuly is based on Ethereum smart contracts and hence do not possess their own blockchain implementation. There are also relatively new projects which combine different technologies to offer video content delivery, sharing, incentivization, security at the same time (e.g., CoinTube \cite{cointube}). As a matter of fact, many of these initiatives can be classified as one of the following combinations as shown in Figure 1. By choosing an open source project for each layer, one can put together a decentralized application (Dapp) and announce an ICO easily if any sort of incentivization is desired. 

\begin{figure}
    \centering
  \includegraphics[width=0.8\linewidth]{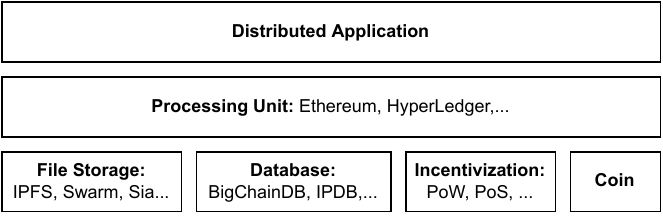}
  \caption{: Layers of Functionality for a decentralized/Incentivized Computer System. PoW: Proof of Work, PoS:Proof of Stake. Coin represents some form of currency used to incentivize the system.}
  \label{fig:boat1}
\end{figure}

Despite all these new technologies centered around open source platforms, today’s technology requirements vary at a great scale as we move from one application to another. For instance, we can note that none of these studies
\begin{itemize}
\item[$a$.] Guarantee the originality of uploaded files, integrity and authenticity of the video content.
\item[$b$.] No verification process for recorded/uploaded videos is explicitly defined. 
\item[$c$.] No supporting proof of time, location, other sensor data to help the verification process of the video authenticity.
\item[$d$.] No genuine immutability (that comprises the full content of data) concept other than the linked list of hashing offered by classical blockchains.
\item[$e$.] No genuine consensus best fitted for video processing/surveillance data and applications. 
\end{itemize}

To address some of these issues, PROVER  project (through ICO) and few later publications (\cite{prover2017}) has recently be crowdsold and attracted attention since this service addressed $a$, $b$ and $c$ to some degree. According to PROVER, mobile device users use Swype ID by moving their cell phone in a specific direction (generated pseudorandomly by the application) to generate code and hashes of the content to be stored in the blockchain. PROVER do not care about where the original content of the data is stored or for some reason whether it is erased. It is particularly designed for checking authenticity and integrity which alone opens up a wide range of applications including video surveillance. However, PROVER is powered by Ethereum or NEM blockchains \cite{nemnetwork} which have their own consensus algorithms predetermined and run by their own development environments (PoW for Ethereum at the time of writing this paper and PoI: Proof of Importance for NEM). In addition, PROVER does treat the video files as a whole and do not use its differentiating features {(such as allowing finer degradation of quality in case of compression, storage or transmission \cite{hammouri2018})} that can be combined with blockchain to provide more efficient and useful recording  experience which will contribute to scalability and flexibility of the overall system. In this study, we will be presenting general architectural components when combined together will best fit in video streaming and surveillance applications. 

{To the best of our knowledge, the closest study in content to our work is VideoChain \cite{liu2018}. However, VideoChain uses a permissioned blockchain i.e., there are trusted parties in the architecture to be able to increase the transaction rates. Their application scenario is based on Campus video surveillance, a rather limited use case. Also, Videochain's blocks are based on video recordings, not on different type of frames as in our study. In VideoChain, the consensus is based on PoS i.e., no proof of work is required, again thanks to the trusted parties in the system. On the other hand, we employ proof of useful work where the usefulness is due to video compression which is required in any multimedia communication scenario. Since the consensus has both useful work and storage component, we believe that the application possibilities of the proposed architecture would be wider.}

The organization of the paper is as follows. In Section 2., we introduce a novel compress-store architecture and provide the details of the proposed system including mining and verification processes. In Section 3, we dive into the details of the implementation and system-level decisions. We also provide advantages and disadvantages of the proposed scheme compared to the state-of-the-art. Finally, section 4 concludes the paper with few future directions.

\section{Compress-Store Architecture}

{In this section, we elaborate on the proposed architecture. In particular, we use} blockchain  for metadata storage (description of which will follow later) while the main contents of the data is stored off-chain using a distributed hash table system. The off-chain choice is completely arbitrary and could be replaced with existing cloud services such as Azure \cite{azure2011} or S3 \cite{amazons3} for as long as they meet the latency and scalability requirements. However, we provide desirable properties of a blockchain applied to bulk data as well such as chaining blocks before moving it to off-chain storage. Here are some attractive features of the proposed compress-store system that distinguishes it from the previous works; \textbf{(a)} Mining/Consensus is based on the novel Proof-of-WorkStore(PoWS) consensus which we will detail later, \textbf{(b)} processing/compression {(main computation framework)} will be decentralized and some compression related parameters will be stored in the blockchain for later verification of recording time, recording place, various sensor information, compression fidelity and {finally,} \textbf{(c)} data is selectively chained and encrypted. We will detail these properties of the system in the next section. 

\subsection{\textit{A Novel Consensus Algorithm: “Proof of WorkStore”}}
Bitcoin’s network uses Proof of Work (PoW) consensus algorithm in which the blocks are mined by solving a mathematical challenge \cite{pow2015}, \cite{nakamoto2008}. This challenge enables the network nodes to reach a consensus and the network in return rewards those nodes who participated in system maintenance and security by offering their CPU resources for solving the challenge. However, for practical use cases which involve video streaming and storage, there are a couple of problems with the original concept of PoW which can be enumerated as follows:
\begin{itemize}
    \item It leads to considerable and useless energy/power consumption. Zero system efficiency results due to finding a solution to a mathematical puzzle that leads to no useful work done. Plus, PoW is typically used to maintain the security of the network, {typically employed} in a public domain. 
    \item The time it takes to show PoW depends on a time-dependent difficulty level. This level increases over time as more miners participate in the ever-growing blockchain network. Increased difficulty level will lead to block mining process to slow down, limiting the scalability of the system. Also in order to limit the chain forking to a minimum, average time between two mining instants is adjusted to meet some criterion. Such adjustments leads to low transaction throughput performance as the size of the block (and hence the number of transactions that it contains) is usually not larger than a predetermined thresholds (1MB in the Bitcoin case - similar sizes apply for other popular public blockchains {such as Ethereum 2.0 \cite{eth20}}). 
    \item In deflationary crypto ecosystems, when mining rewards cease, only transcation fees will incentivize the system. Once these fees drop, the number of miners will decline for service leading to unsecure and unprotected system design. 
\end{itemize}

The other alternative applicable consensus methods include Proof of Stake (PoS) \cite{eth20}, Proof of Space (PoSpace), Proof of Storage (PoSt) and Proof of Importance (PoI) etc \cite{sankar2017}. We immediately realize that none of these methods require elevated CPU and ASIC requirements for better throughput performance leading to greener decentralization. In PoSt, miners have to show a proof for enough storage space to store the corresponding data and will have to guarantee that it never erases data in their local or remotely owned storage slots. There are few ways of implementing PoSt in literature depending on what is exactly being achieved. Some of PoSt schemes include Proof of Retrievability (PoR) \cite{shacham2008}, Provable Data Possession (PDP) \cite{liPOS2020}, Proof of Replication (PoRep) \cite{Benet2017} and the most common implementations to all is to use cryptographic operations and periodic auditing protocols \cite{atheniese2007}. In a typical application, the provers generate a set of challenges which requires accesses to random parts of the data. {T}he generations of such proofs are performed at random times with a limitation on the time between two consecutive challenges generated for the same data.

Since the ideal decentralized computer system is expected to establish both decentralized computing and storage at the same time, it is essential that we provide in our video surveillance system \textbf{(1)} Video Processing (compression in our particular case) \textbf{(2)} Data storage \textbf{(3)} Immutability and \textbf{(4)} Security all at the same time. Note that providing such qualities in a conventional centralized framework would lead to inefficient utilization of resources, governance of one body or organization in our surveillance application {(centralization)} and increased system costs.

In our context, video compression will have two {important} advantages. First it requires some form of computation due to video processing (transformation, quantization and subsequent entropy coding) and this can be done in a decentralized fashion and yet does not pose a lot of useless computations as in original PoW and the difficulty of \textit{the challenge} only changes as new compression algorithms and techniques or new quality requirements are integrated/imposed into/onto the network. This will form the proof of work part of our consensus, namely PoWS. The second advantage is that since raw video files would be compressed at a specified quality, we can save a lot of storage space (efficient utilization of data storage resources). This does not only mean that we will be saving storage resources but also computation resources that might be due to encryption, digital signature generation etc. Note that one other advantage comes at no cost from the incentivization point of view, because miners may want to choose high compression performance to be able to find storage place quickly and hence be successful at their mining process. Through such incentivization, it will lead to total storage of the system to be used more wisely.  

Video files are bulky and it is always hard to deal with large volumes of data in general. In the compress-store architecture, the data storage is provided off-chain using a distributed hash table system. One possible realization is the content-addressed data chunk storage technologies such as IPFS. In that case the data location is represented by a unique hash and we separate the content location in the network and the IP/port number of the server. However, all location pointers will be stored in the blockchain. Finally since the metadata is stored in the blockchain, it is immutable and cannot be changed by any easy means. 

The main idea behind inserting some kind of smarter PoW into our system is  to first dramatically increase the scalability of the system and it would make really hard to generate compressed sequences of thousands of frames in a relatively short period of time (unless application specific hardware is used. However, keep in mind that video encoders comes in great variety of parameter selections and algorithmic differences) which discourages attackers and allows the network to use the proposed system in the public domain. 

\subsection{\textit{Throughput of the Surveillance System}}

%In one application of the proposed idea, miners compress video frames. 
Video files are typically partitioned into Group of Pictures (GOP) and each is processed independent of the other. In other words, each GOP is treated as the smallest unit subject to processing and processing of GOPs can be concurent if processed by different miner nodes. Hence we can define throughput to be the processed frames committed to the blockchain per second. In a typical scenario, a GOP can contain 25 frames and each block can contain around 5 GOPs at the same time. In an optimistic scenario, if each block is verified and added to blockchain every 10 secs, this would make 12.5 frames per second (pps). This is an extremely slow rate compared to the level of video generation by the system in a typical surveillance application. There are multiple ways of improving the throughput of the system. One obvious way to improve the throughput is to increase GOP size at the expense of lesser quality compression and larger storage requirement. Number of nodes to store the compressed content will be a performance limiter. Another popular way to alleviate this is the method of sharding \cite{sharding2016} that is also being considered to solve the scalability issues of popular public blockchains such as Ethereum. In that scheme, miners choose a GOP or a consecutive group of GOPs pseudorandomly and work on their compression workload. This would allow parallelism in the network and hence would ensure better throughput performance. However if the storage is offloaded, accessing such storage nodes and committing the compressed content will determine the final throughput performance. On the other hand, implementation of such a scheme might be a bit tricky. 

Note that due to twin cosensus used by the system and dependence on the network bandwidth, the variance of the throughput of the system is expected to be high. Thus, another and more secure approach would be to make the blockchain private. In that case, the security requirements will be less of an issue due to trusted parties, communication links, routing and the traffic on such links would be more controlled and hence more frequent block additions to the chain could be realized. This would eventually lead to better throughput. 

\subsection{\textit{Private/Public Blockchain applications}}

In a private blockchain implementation of the proposed idea, compression workload can completely be handled by the video generator node also called as \textit{initiator node}. This process can alternatively be handled completely decentralized manner if need be. In that case, the PoW part of our system can be avoided since the participants are assumed to be trusted parties and pose no risk to the system. This way, the number of committed video frames into the blockchain can be increased dramatically and consensus can be reached a lot easily. This will eventually increase the throughput of the system. Although a form of centralization may dominate (also due to a subset of miner node selection process), all other properties of the proposed scheme will still serve a number of advantages regarding the video surveillance applications. 

In one public blockchain implementation of the proposed idea, we shall use PoWS at full scale. Compared to private counterpart, there are number of differences in this case. First, we decentralize the computation by allowing miner nodes to compress and encrypt video frames, find an appropriate storage location before preparing and adding the related metadata into the blockchain. Since these are third party participants, we propose to incentivize them by coins which will help them process more videos and use storage space. We also incentivize better compression (avoid dump compression styles) because finding an appropriate storage location and space can only be found through paying the required amount using coins. Miner nodes are motivated to use better methods to be able to ask for less storage space that also meets a predefined (just like difficulty level of a Bitcoin network) quality requirement. This quality requirement may be updated as the network evolves or more miners participate. We refer the reader to the mining process for details. 

\section{Implementation and System Details}

%In this section, we provide the implementation and system-level details. some of these details are related to verification and data storage phases. Later, we shall compare these details with the state-of-the-art. 

\subsection{\textit{The procedure for Verification and Storage}}

We define three sorts of nodes in our compress-store architecture, each with its own set of behaviors, as shown below.
\begin{itemize}
    \item[1.] \textbf{Initiator nodes:} These type of nodes are usually equipped with video camcorders and are able to capture, record the raw data and modify/edit the recorded video streams. These streams can be divided into one or more GOPs in an edit mode. The captured GOPs are usually selected to be small and they constitute the block of information to be processed by the network.
    \item[2.] \textbf{Mining nodes:} These nodes are equipped with video compression tools/encoding software or hardware and ensures the security of the system throughout the operation. They process received frames of a GOP and send identification information (such as to show the intended compression is performed) to the network to initiate the verification process. Once their work is verified, a block representing one GOP is added to the blockchain and the mining nodes/addresses are rewarded with fees.
    \item[3.] \textbf{Storage nodes:} These nodes are responsible of storing bulk data which is in our case the compressed multimedia source files. These files are stored encrypted after compression. Storage nodes run a form of PoSpace algorithm to make sure that they reserved the amount of space that they promise to. Although block verification verify the availability of data and the storage space, a travelling auditing service shall be used by the network to check this verification process at a regular basis. Storage nodes receive fees (or a cryptocurrency as the incentivisation mechanism) once they complete all the requirements of the PoSt and as long as they store the multimedia data. 
\end{itemize}

Physical nodes of our peer-2-peer network can assume all these three types of node capabilities. For instance, \textit{full nodes} can initiate video storage, mine frames and store compressed multimedia data all at the same time. On the other hand, \textit{verifier} nodes are all  participating network nodes that store a copy of the blockchain and take on the responsibility of verifying both the compression as well as storage works done by the miners and storage nodes, respectively. 

\begin{figure}
\centering
  \includegraphics[width=0.91\linewidth,height=5.7cm]{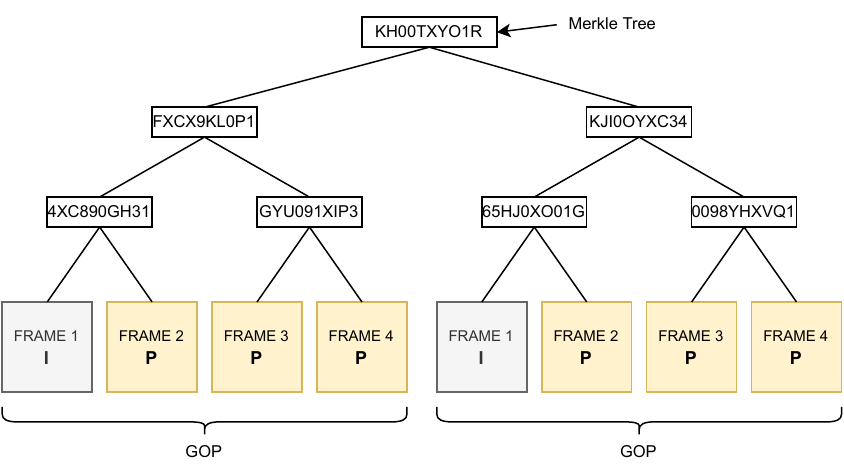}
  \caption{An example of two GOPs and the procedure of Merkle root computation within and across GOPs. In other words, merkel roots of each GOP are combined to compute one common merket root that stays at the top of the hierarchy.}
  \label{fig:merkle_tree}
\end{figure}

The multimedia data itself is NOT stored in the blockchain. Instead their representative data is maintained on-chain serving as pointers to their stored locations. We define a subblock to include the following information: 1. The time GOP is generated, 2. Hash of the GOP (through Merkle root), 3. Access privileges, 4. The location of stored GOP (this does not need to be a physical address, an alternative is to use content addressing, also geographical location can be incorporated), 5. Metadata about the stored multimedia content such as compression algorithm, parameters, 6. transactions regarding the awards for executing compression and data storage operations, etc. In addition,
variety of sensor information, GOP index and order, video identification number/labeling
could also be part of the subblock for further verification process. These additional information is important for the reconstruction of the video files. Once a subblock (with 1-5) is formed it is digitally signed with initiators’ private key before sending in to network. Miner nodes shall collect enough number of such subblocks and the associated multimedia data frames/GOPs to start compression process immediately after such subblocks make up a predetermined block size (determined by the overall network). This predetermined block size is analogous to the block size of other crypto-networks such as Bitcoin. If miner nodes do not store compressed video themselves then they are required to find out storage nodes to store compressed content and their addresses. After securing storage space and commit, they insert necessary metadata as well as transactions (6) into the subblocks for further verification.

For compression to make sense, an initiator must set a quality measure such as Mean Square Error (MSE) or a Peak Signal to Noise Ratio (PSNR) or an another subjective multimedia quality indicator that will help identify that the original files and the compressed version are the same visually subject to a quality measure. There are few technologies/algorithms that can differnetiate two video files whether they are compressed or not. Miners’ time to mine a block requires the miner to complete the compression process (by going through each sub-compression steps such as transformation, quantization and entropy coding etc.), meet the quality measure requirement, compute the Merkle hash tree of a GOP, chain the video frames, encrypt the content, find a storage node (or uses his own local resources) which ensures storage space required to store the compressed content (generate a proof). In order to make the size of a block even smaller, we employ Merkle hash tree of GOPs as well. Hence, multiple trees can be combined to make up a second layer of Merkle hash tree. The way the common Merkle root is computed is shown in Figure 2. 

\begin{figure}[t!]
\centering
  \includegraphics[width=\linewidth]{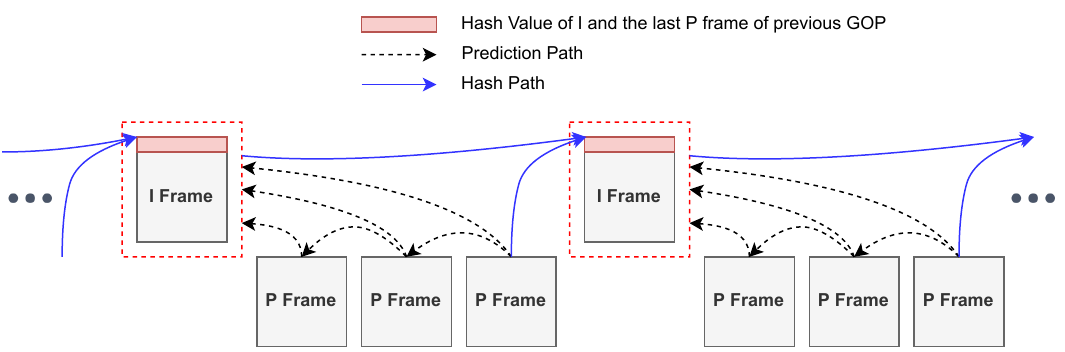}
  \caption{The idea of data chaining combined with video compression and predictive frame coding. The hash value of the previous block/s can also be included.}
  \label{fig:data_chaining}
  \vspace{-2mm}
\end{figure}

Once this work is done, the mining node broadcasts the block/s to the network that has all the information about the GOP except the raw data itself. All verifier nodes which receive this block begin the verification process. The verification process involves:
\begin{itemize}
    \item[a.] A comprehensive check whether GOPs are really stored in the designated locations. This requires preparing intelligent challenges for provers (miners) for PoSt. For instance, hash values of randomly selected parts of the compressed data may be requested. In case the storage nodes could not provide that information on time, their fees are not paid. This is accomplished through either cancelling the corresponding transactions or adding transactions to undo the previous payments.
    \item[b.] A comprehensive check for the merkle hash by requesting hashes of the frames and GOPs from storage node/s. a seperate Merkle tree root check is also conducted for the transactions.
    \item[c.] A comprehensive check for the quality measure whether the compression work meets it or not, using the uncompressed GOP data. This is to ensure that the miner nodes are legitimately compressed and stored the compressed multimedia file. This effort in fact characterizes a form of Proof of Compression (PoC). 
\end{itemize}

Once verified, all uncompressed copies are removed from verifier node caches to open space for the next uncompressed GOP/s. We keep the size of the blocks that contain metadata for GOP/s to around only a fraction of KBs. This is to limit the total size of the blockchain stored in all of the verifier nodes.  One can realize that as we include more metadata (transactions, sensor information etc.) in the blockchain, we can increase the security, authenticity and reliability of the system at the expense of reduced scalability and throughput, which is infact the fundamental trade-off any blockchain system faces today. 

Depending on the compression scheme, video frames can be predicted. In a typical compression scenario, we can classify frames as I and P where I is intra-coded frame i.e., the image gets compressed all by itself whereas P frames are predicted from the associated I frame and one previous P frame. In addition, we can have B frames that shall be predicted from two or more P frames. A GOP will contain one I frame in the beginning and all the rest would be P (and/or B) frames. In order to chain the data for immutability, I frames of a video source file are selected to contain a hash value of the previous I frame and the latest P frame in the previous GOP. Due to predictive nature of compression algorithm, P frames are automatically chained to I frames and hence are not separately chained using cryptographic functions. This will reduce the computation requirements due to hashing. A detailed illustration of how the prediction and the hashing are done all together in the compress-store architecture is briefly shown in Fig. 3. Let us provide the summary of steps for a full node to initiate, mine, store and verify a recorded video. 

\begin{figure}[t!]
\centering
  \includegraphics[width=0.85\linewidth]{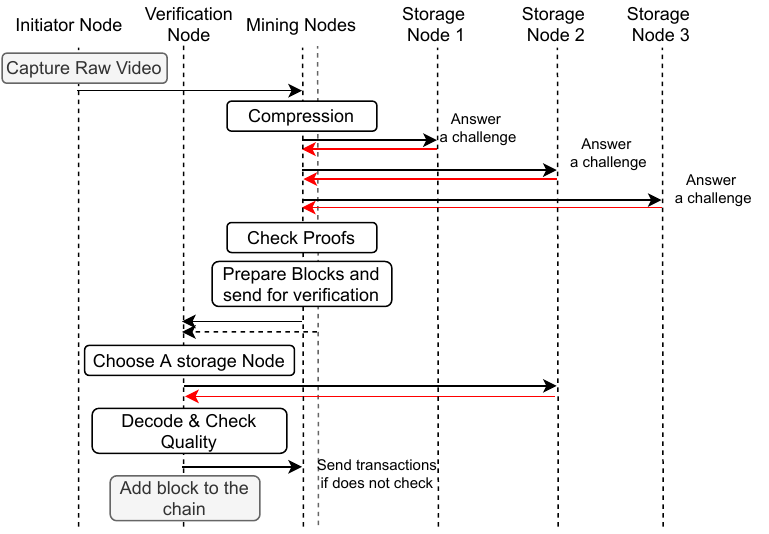}
  \caption{Initialization, mining and verification processes are all subject to the availability of enough storage space and computation power. Miners select multiple storage nodes to maintain availability and prevent compressed data loss. Challenge questions is at the heart of PoSt besides PoW. We used different colors to indicate requests and responses. Alternative methods such as erasure coding can be used in place of replication to save storage space.}
  \label{fig:operation}
  \vspace{-1mm}
\end{figure}

\begin{enumerate}
    \item[(1)] After video file is captured, it is streamed/broadcasted to the miner network nodes GOP by GOP (GOPs can be thought as image transactions in crypto context and to be able to differentiate them from real transactions these are referred as GOP transactions (gtxns) in our context) where each gtxns  is digitally signed by the issuer for authentication. This step is used to prevent potential outsourcing and authenticate the work (both for PoW and PoSt). The specific format of GOP transactions are implementation-specific. If GOP transmission overwhelm the available network, then the content address of these GOPs can be shared instead and the participating mining and verifier nodes can download them for processing{, paying both for network bandwidth and CPU clock time}.
    \item[(2)] Miner nodes collect/pack a set of gtxns, authenticate them, process (compress) them and then compose a (associated) set of subblocks to make up a block (block size determines the number of subblocks that can be bundled together) that also contains the hashes (Merkle roots) of the previous block. 
    \item[(3)] In an application, miners’ PoW may include compression and encryption of the set of collected/packed gtxns. Real transactions could be part of the PoW directly (using the standard nonce calculation etc.) or can become part of compression and encoded into the blockchain through compression and encryption. One such possibility is to hide transaction data inside the video. The way we insert this data into the GOP can be done randomly and hence miners can compress the same GOP multiple times and send the one with the highest quality.
    \item[(4)] On the other hand, miners’ PoSt include a proof for the storage of the compressed and encrypted content included with the prepared block. In an attempt to replicate the stored compressed content, the miner shares the final processed GOP with multiple storage nodes. The storage can be provided with cloud services or any other peer of the network with local persistent storage. Miner uses a challenge question to check whether the storage nodes store the final compressed content. On the other hand, with digital signature requirement, outsourcing may be forbidden or unincentivised (such as reducing the fees earned) on purpose by the internal system management. If not forbidden, then It is typical that if the miner and storage node are the same physical node, then fees paid may be higher to make storage offloading less attractive.
    \item[(5)] After storage, miners place the location of the compressed data (content address), associated hash values such as merkle tree root of gtxns or txns or leaf node hashes, compression algorithm name and parameters, quality measure, any additional data (such as proofs) that would be useful for verification into the (subblocks) blocks and broadcast it for verification. 
    \item[(6)] Verification nodes (also referred as verifiers) read the contents of the block and easily verify whether the content is accurately compressed, properly encrypted, and stored by contacting the appropriate storage nodes, according to a predefined quality measure (using the proofs included with the blocks). 
    \item[(7)] Once the verification process successfully ends, the block is added to the local blockchain. If more than one miner's sublock arrives, and all of them checks, then the one with the highest quality metric is added to the blockchain. Also, if any one of the requirements is not satisfied, the block is not added to the blockchain  and the verifier node moves on to the next verification process waiting in the network. 
    \item[(8)] Additionally, verifier nodes can also create challenge questions based on the previous GOP compression works which are present in the blockchain.
    In case, one of the storage nodes fail to show an evidence, then a corresponding transaction is generated and send to miners for further processing. This way fees already assigned to storage nodes can be deducted for not complying with the {protocol} of the system.
\end{enumerate}

\begin{figure}[t!]
\centering
  \includegraphics[width=0.7\linewidth]{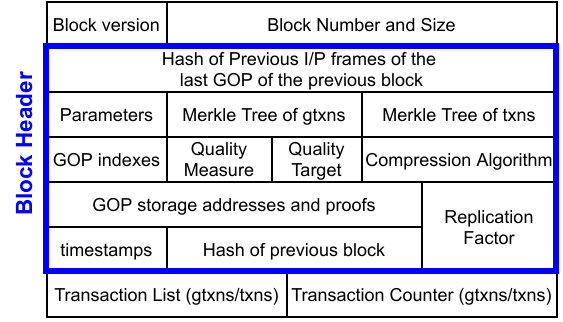}
  \caption{The block structure and format is heavily dependent on the use of the techniques mentioned in the text as well as the implementation details. Here we provide a sample block format that can be used to implement the proposed scheme. The number of bits used for each field is again a function of the techniques used and implementation requirements.}
  \label{fig:blockinternals}
  \vspace{-1mm}
\end{figure}

A representative functional diagram is depicted in Fig. \ref{fig:operation} to picture the details of the operation. 

{
\subsection{Analysis of the Proposed system}}

{
First of all, the correctness analysis of the system is quite close to that of \cite{liu2018}. In other words, each verifier can authenticate the proof on their own, and no initiator node can unilaterally delete the encoded and stored video. Moreover, due to PoSt, proofs generated based on the challenge questions also ensures the integrity of the data stored. Although due to compression, encoding the integrity into the blockchain may be slow, however, ordering based on the quality of the compression work ensures only one of the mining jobs to be mined into the blockchain and hence guarantee of convergence. The proposed scheme is also resilient against Sybil attacks. This is due to the adversary should control more than 50\% of all the nodes in blockchain. This means that one needs to possess more than 50\% of overall compression and storage resources in the system to violate all the guarantees, which is almost impossible to achieve.  Finally, all compression work and decisions made on the storage space are traceable and temper-proof due to the characteristics of the blockchain and hashing involved in the production of the blocks in the proposed scheme.}

\vspace{4mm}

\subsection{\textit{Some potential problems and workarounds}}
One of the risks is about the bandwidth requirements of the proposed system as the system heavily deals with bulky data communications. Particularly in the public domain, broadcasting the whole raw video, even GOP by GOP might be too bandwidth consuming and will lead to data traffic. Most of this traffic is useless (except for the node that successfully fulfill all the requirements of mining). As a solution,	initiator nodes may store their raw videos either locally or remotely before broadcasting the location and content address--hash value of  data for the miners to download and check later. This will make download speed and bandwidth be part of the equation in our PoWS definition. In case of sharding, miners may only download GOPs that are not mined yet which will lead to efficient use of network resources as well. 

The other important issue is known as the forking of the blockchain which comes out due to the multiple copies of the blockchain that result due to concurrent verification processes. As the verifiers connect to multiple and potentially distinct miners, and since transaction data introduces a form of randomness in the compression framework, the the way the instructions are included in the compression process are quite different, the quality of the compression would be different. Therefore, as the verifiers add blocks to the blockchain, they can keep track of the accumulated quality of the compressed video and in the case of forking, through an exchange, they adapt the one with the highest quality in the network. Consensus is achieved as the blockchain with the highest quality score is adapted by all the nodes in the network.

\subsection{\textit{Advantages compared to the state-of-the-art}}

First of all, the proposed scheme ensures data immutability through the use of a blockchain as well as data chaining process that connects $I$ frames (self-compressed frames) of the compressed video. The predictive nature of the compression process is used to add an extra layer of chaining between different kinds of compressed video frames in addition to hash chaining of the blockchain. Thus, any tempering on data can immediately be detected since this attempt will change all hash outputs in a propagated fashion. Plus, changing a block content will require all the following blocks to change which will require PoWS for all GOPs, represented by these blocks. This would require a large volume of specific computation as well as large amounts of secured storage space. Secondly, the concept of PoWS does not allow our system to have a solely CPU-based mining which can lead to hardware specific implementations and hence centralization. Additionally unlike bitcoin’s consensus mechanism, PoW component of PoWS requires miners to perform useful work, i.e., in an application of the proposed idea, a miner can compress a video file using his CPU and network resources. This way, miner will make his job easier when finding external nodes to store the compressed content. More sophisticated compression will help miners spend less for data storage and bandwidth at the expense of more CPU power. On the other end, Miners can choose to go with a simpler compression technique at the expense of larger storage space committed for the compressed content. Such variations of the proposed idea lead miners to complete the total work of PoWS at different instants of time and hence block generation happens at relatively different times. As a result of that, potential (soft) forks (in the blockchain convergence process) will be eliminated without getting too lengthy i.e., convergence of consensus will be faster.  

\vspace{-1mm}
\section{Conclusion and Future work}
In this study, a genuine video surveillance system based on DLTs is presented. the proposed scheme uses data compression and storage as means of proofs in order to provide green consensus to help network participants to commit their resources for useful work. Finally, the details of the proposed scheme is presented by providing comparisons to state-of-the-art schemes {along with some theoretical evidences}. As a future work, we would like to extend the implementation details of the system to include image files. The application areas can be extended to include image restoration, multimedia regeneration and data mining. Finally, large-scale simulations are envisioned to demonstrate the performance and effectiveness of the proposed idea in real test beds.

%\begin{acknowledgements}
%If you'd like to thank anyone, place your comments here
%and remove the percent signs.
%\end{acknowledgements}

% Authors must disclose all relationships or interests that 
% could have direct or potential influence or impart bias on 
% the work: 
%
% \section*{Conflict of interest}
%
% The authors declare that they have no conflict of interest.

% BibTeX users please use one of
%\bibliographystyle{spbasic}      % basic style, author-year citations
%\bibliographystyle{spmpsci}      % mathematics and physical sciences
%\bibliographystyle{spphys}       % APS-like style for physics
%\bibliography{}   % name your BibTeX data base

\begin{thebibliography}{}

\bibitem{Nordrum2016} Nordrum, A. (2016). The internet of fewer things [news]. IEEE Spectrum, 53(10), 12-13.

\bibitem{amazons3} Palankar, M. R., Iamnitchi, A., Ripeanu, M., \& Garfinkel, S. (2008, June). Amazon S3 for science grids: a viable solution?. In Proceedings of the 2008 international workshop on Data-aware distributed computing (pp. 55-64). ACM.

\bibitem{pilkington2016} Pilkington, M. (2016). 11 Blockchain technology: principles and applications. Research handbook on digital transformations, 225.

\bibitem{lamport2001} Lamport, L. (2001). Paxos made simple. ACM Sigact News, 32(4), 18-25.


\bibitem{raft2014} Ongaro, D., and Ousterhout, J. (2014). In search of an understandable consensus algorithm. In 2014 {USENIX} Annual Technical Conference ({USENIX}{ATC} 14) (pp. 305-319).

\bibitem{baliga2017} Baliga, A. (2017). Understanding blockchain consensus models. In Persistent.

\bibitem{mermer2018} G. B. Mermer, E. Zeydan and S. S. Arslan, (2018) An overview of blockchain technologies: Principles, opportunities and challenges, 2018 26th Signal Processing and Communications Applications Conference (SIU), pp. 1-4.

\bibitem{nakamoto2008} Nakamoto, S. (2008). Bitcoin: A peer-to-peer electronic cash system.

\bibitem{wood2014} Wood, G. (2014). Ethereum: A secure decentralised generalised transaction ledger. Ethereum project yellow paper, 151, 1-32.

\bibitem{tseng2020}  L. Tseng, X. Yao, S. Otoum, M. Aloqaily, and Y. Jararweh, ``Blockchainbased database in an iot environment: challenges, opportunities, and analysis,” Cluster Computing, pp. 1–15, 2020.

\bibitem{arslaniot2020} Arslan, S. S., R. Jurdak, J. Jelitto, and B. Krishnamachari. 2020. “Advancements in Distributed Ledger Technology for Internet of Things.” Internet of Things 9 (1): 100114.

\bibitem{Karafiloski2017} Karafiloski, E., and Mishev, A. (2017, July). Blockchain solutions for big data challenges: A literature review. In IEEE EUROCON 2017-17th International Conference on Smart Technologies (pp. 763-768). IEEE.

\bibitem{bittorent2005} Pouwelse, J., Garbacki, P., Epema, D., and Sips, H. (2005, February). The bittorrent p2p file-sharing system: Measurements and analysis. In International Workshop on Peer-to-Peer Systems (pp. 205-216). Springer, Berlin, Heidelberg.

\bibitem{ipfs2004} Benet, J. (2014). Ipfs-content addressed, versioned, p2p file system. arXiv preprint arXiv:1407.3561.

\bibitem{maymounkov2022} P. Maymounkov and D. Mazieres, ``Kademlia: A peer-to-peer
information system based on the xor metric,” in International
Workshop on Peer-to-Peer Systems. Springer, 2002, pp. 53–65.

\bibitem{nazadeh2021} Hassanzadeh-Nazarabadi, Y., Taheri-Boshrooyeh, S., Otoum, S., Ucar, S., \& Ozkasap, O. (2021). DHT-based Communications Survey: Architectures and Use Cases. arXiv preprint arXiv:2109.10787.

\bibitem{drago2012} Drago, I., Mellia, M., M Munafo, M., Sperotto, A., Sadre, R., \& Pras, A. (2012, November). Inside dropbox: understanding personal cloud storage services. In Proceedings of the 2012 Internet Measurement Conference (pp. 481-494). ACM.

\bibitem{vorick2014} Vorick, D., \& Champine, L. (2014). Sia: Simple decentralized storage. White paper available at https://sia. tech/sia. pdf.


\bibitem{storj2014} Wilkinson, S., Boshevski, T., Brandoff, J., and Buterin, V. (2014). Storj a peer-to-peer cloud storage network.

\bibitem{hartman1999} J. H. Hartman, I. Murdock, and T. Spalink, “The swarm scalable storage system,” in Distributed Computing Systems, 1999. Proceedings. 19th IEEE
International Conference on. IEEE, 1999, pp. 74–81.

\bibitem{filecoin2014} Techical Report. Filecoin: A Cryptocurrency Operated File Network. http:
//filecoin.io/filecoin.pdf, 2014.

\bibitem{maidsafe2014} Paul, G., Hutchison, F., \& Irvine, J. (2014, May). Security of the MaidSafe vault network. In Wireless World Research Forum Meeting 32 (WWRF32).

\bibitem{shafagh2017} Shafagh, H., Burkhalter, L., Hithnawi, A., and Duquennoy, S. (2017, November). Towards blockchain-based auditable storage and sharing of iot data. In Proceedings of the 2017 on Cloud Computing Security Workshop (pp. 45-50). ACM.

\bibitem{bigchaindb2016} McConaghy, T., Marques, R., Müller, A., De Jonghe, D., McConaghy, T., McMullen, G., \& Granzotto, A. (2016). BigchainDB: a scalable blockchain database. white paper, BigChainDB.

\bibitem{stampery2015} Dillet, R. (2015). Stampery Now Lets You Certify Documents Using the Blockchain and Your Real Identity. Nov, 20, 6.

\bibitem{verify2018} Verif-y: Blockchain Solution For Sustainable Self-Sovereign Identity. Available online; https://verif-y.com/.

\bibitem{livepeer2017} Teutsch, J., \& Reitwießner, C. (2017). A scalable verification solution for blockchains. URL: https://people. cs. uchicago. edu/teutsch/papers/truebit pdf.

\bibitem{liu2016} Liu, M., Teng, Y., Leung, V. C., \& Song, M. A Novel Resource Management Scheme for Blockchain-based Video Streaming with Mobile Edge Computing.

\bibitem{cointube} COintube: Decentralized Video Platform. Available online: https://cointube.org/.

\bibitem{prover2017} Liang, X., Shetty, S., Tosh, D., Kamhoua, C., Kwiat, K., \& Njilla, L. (2017, May). Provchain: A blockchain-based data provenance architecture in cloud environment with enhanced privacy and availability. In Proceedings of the 17th IEEE/ACM international symposium on cluster, cloud and grid computing (pp. 468-477). IEEE Press.

\bibitem{nemnetwork} NEM Technical Reference, Version 1.2. 2018. [Online]. Available online: https://nem.io/wp-content/themes/nem/files/NEM\_tech.

\bibitem{hammouri2018} M. Al-hammouri, B. Madani, M. Aloqaily, I. A. Ridhawi and Y. Jararweh, "Scalable Video Streaming for Real-Time Multimedia Applications over DDS Middleware for Future Internet Architecture," 2018 IEEE/ACS 15th International Conference on Computer Systems and Applications (AICCSA), 2018, pp. 1-6.

\bibitem{liu2018} M. Liu, J. Shang, P. Liu, Y. Shi, and M. Wang, ``VideoChain: Trusted video surveillance based on blockchain for campus,’’ in Cloud Computing and Security—ICCS (Lecture Notes in Computer Science), vol. 11066, X. Sun,
Z. Pan, and E. Bertino, Eds. Cham, Switzerland: Springer, 2018, pp. 48–58.


\bibitem{azure2011} Calder, B., Wang, J., Ogus, A., Nilakantan, N., Skjolsvold, A., McKelvie, S., \& Haridas, J. (2011, October). Windows Azure Storage: a highly available cloud storage service with strong consistency. In Proceedings of the Twenty-Third ACM Symposium on Operating Systems Principles (pp. 143-157). ACM.

\bibitem{pow2015} Vukolić, M. (2015, October). The quest for scalable blockchain fabric: Proof-of-work vs. BFT replication. In International workshop on open problems in network security (pp. 112-125). Springer, Cham.

\bibitem{eth20} “Ethereum 2.0 phases,” https://docs.ethhub.io/ethereum-roadmap/
ethereum-2.0/eth-2.0-phases/, 2019.

\bibitem{sankar2017} Sankar, L. S., Sindhu, M., \& Sethumadhavan, M. (2017, January). Survey of consensus protocols on blockchain applications. In 2017 4th International Conference on Advanced Computing and Communication Systems (ICACCS) (pp. 1-5). IEEE.

\bibitem{shacham2008} Shacham, H., and Waters, B. (2008, December). Compact proofs of retrievability. In International conference on the theory and application of cryptology and information security (pp. 90-107). Springer, Berlin, Heidelberg.

\bibitem{liPOS2020} Li, Y., Yu, Y., Chen, R., Du, X., and Guizani, M. (2020). IntegrityChain: provable data possession for decentralized storage. IEEE Journal on Selected Areas in Communications, 38(6), 1205-1217.

\bibitem{Benet2017} Benet, J., Dalrymple, D. and Greco, N. (2017). Proof of replication. Protocol Labs Technical Report, July, 27, 20.


\bibitem{atheniese2007} Ateniese, G., Burns, R., Curtmola, R., Herring, J., Kissner, L., Peterson, Z., \& Song, D. (2007, October). Provable data possession at untrusted stores. In Proceedings of the 14th ACM conference on Computer and communications security (pp. 598-609). ACM.

\bibitem{sharding2016} Luu, L., Narayanan, V., Zheng, C., Baweja, K., Gilbert, S., \& Saxena, P. (2016, October). A secure sharding protocol for open blockchains. In Proceedings of the 2016 ACM SIGSAC Conference on Computer and Communications Security (pp. 17-30). ACM.
\end{thebibliography}

% Non-BibTeX users please use

\section*{Author Biographies}

\begin{wrapfigure}{l}{25mm} 
    \includegraphics[width=1.1in,height=1.4in,clip,keepaspectratio]{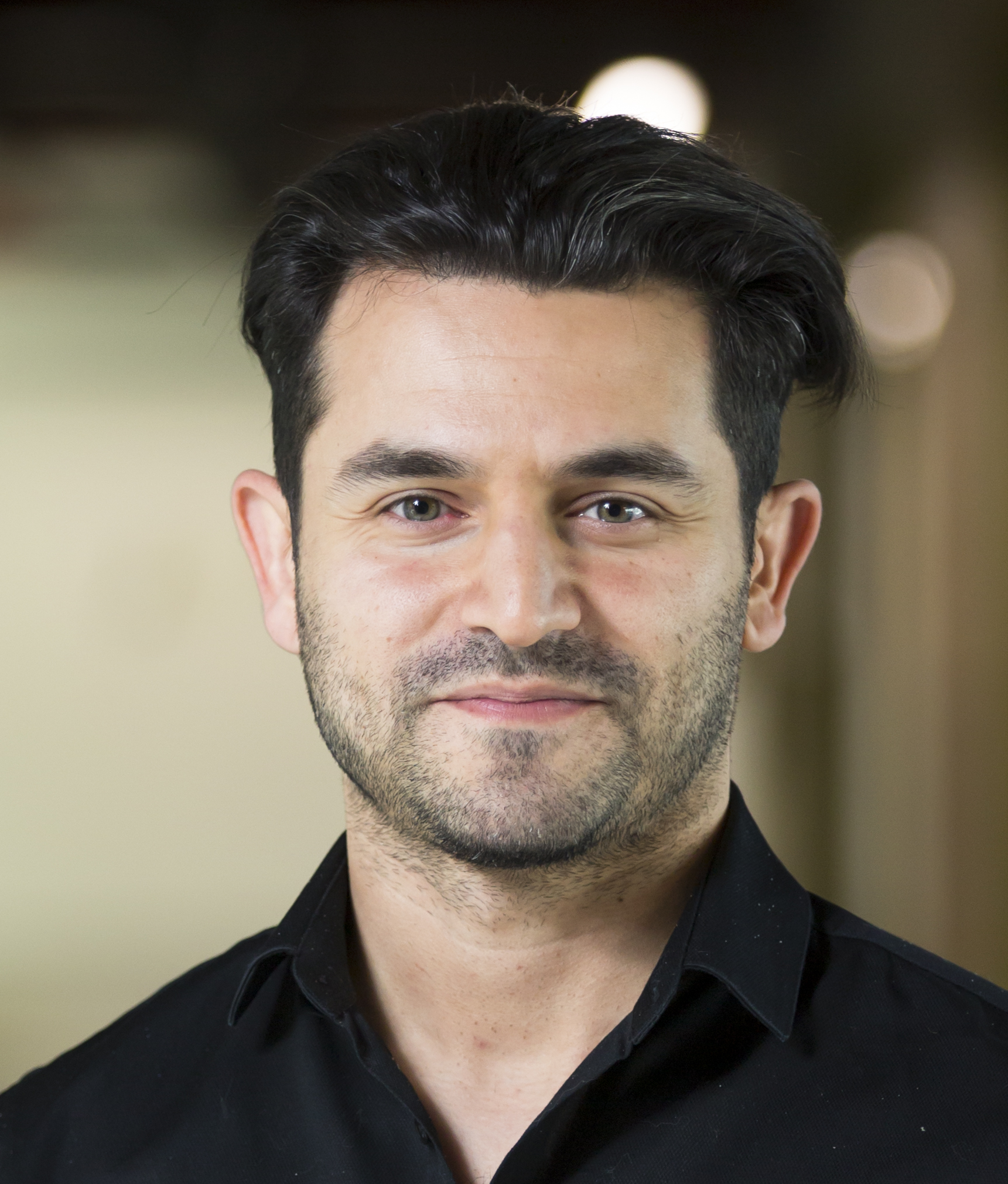}
  \end{wrapfigure} \par
  \textbf{Suayb S. Arslan} Suayb S. Arslan received the B.Sc. degree in electrical and electronics engineering from Bogazici University, Istanbul, Turkey, in 2006, and the M.Sc. and Ph.D. degrees in electrical engineering from the University of California, San Diego, CA, USA, in 2009 and 2012, respectively. He was with Mitsubishi Electric Research Laboratory, Boston, MA, USA, in 2009, where he was involved in research and development of image and video processing algorithms for biomedical applications. In 2011, he joined Quantum Corporation, Irvine, CA, USA, where he conducted research on advanced detection and coding algorithms for increased capacity Tape storage and cloud systems. He is currently affiliated with MEF University as an associate professor. He serves as vice-chair for IEEE ComSoc Turkey and Treasurer for IEEE Data Storage Technical committee. His research interests include digital communication and storage, cloud and Quantum computing, information and reliability theory, image/video processing, and cross-layer design optimizations at the edge. \par
 \vspace{5mm}
 
 \begin{wrapfigure}{l}{25mm}  
      \includegraphics[width=1.4in,height=1.2in,clip,keepaspectratio]{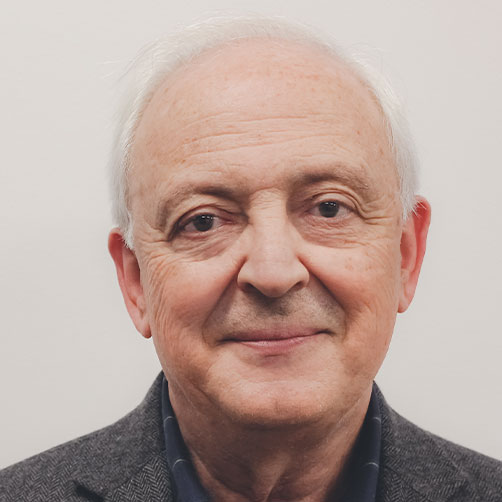}
  \end{wrapfigure}\par
  \textbf{Turguy Goker} is a technologist and the manager of adv. dev. lab. in Quantum Corporation, Irvine, CA. He is an active member of INSIC, SNIA and few other data storage consortiums. He has extensive experience in data storage business. He wrote more than 10 papers in peer-reviewed journals and owns more than 60 patents related to cold data storage, erasure coding, LTO format and distributed databases. \par

\end{document}